\documentclass{article}

\usepackage{titling}
\title{SLO-ML: A Language for Service Level Objective Modelling in Multi-cloud Applications}

\usepackage[sc]{mathpazo} 
\usepackage[T1]{fontenc} 
\usepackage{microtype} 
\usepackage[hmarginratio=1:1,top=25mm,columnsep=20pt]{geometry} 
\usepackage[hang, small,labelfont=bf,up,textfont=it,up]{caption} 
\usepackage{booktabs}
\newcommand{\midsepremove}{\aboverulesep=0mm \belowrulesep=0mm}
\midsepremove
\newcommand{\midsepdefault}{\aboverulesep=0.5mm \belowrulesep=0.15mm}
\midsepdefault
\usepackage{makecell,tabularx,ragged2e}
\usepackage{float} 
\usepackage{paralist} 

\usepackage{abstract} 

\usepackage{titlesec} 
\renewcommand\thesubsection{\Roman{subsection}} 
\titleformat{\section}[block]{\large\scshape\centering}{\thesection.}{1em}{} 
\titleformat{\subsection}[block]{\large}{\thesubsection.}{1em}{} 

\usepackage{fancyhdr} 
\usepackage{url}

\usepackage{amsfonts,amsmath,amssymb,amstext,latexsym}
\usepackage{array}
\usepackage{color}
\usepackage[usenames,dvipsnames,table]{xcolor}
\usepackage{multirow}

\usepackage{graphicx}
\usepackage[caption=false]{subfig}
\newcommand{\fig}[1]{Fig.~\ref{#1}}

\newcommand*{\sect}[1]{\S\ref{#1}}
\usepackage{xspace}

\newcommand*{\eg}{\textit{e.g.}\@\xspace}
\newcommand*{\ie}{\textit{i.e.}\@\xspace}
\makeatletter
\newcommand*{\etc}{%
    \@ifnextchar{.}%
        {etc}%
        {etc.\@\xspace}%
}
\newcommand*{\etal}{%
    \@ifnextchar{.}%
        {et al}%
        {et al.\@\xspace}%
}
\makeatother

\newcommand{\name}{\texttt{SLO-ML}\@\xspace}

\usepackage{pifont}

\usepackage{listings}
\lstdefinelanguage{XML}
{
  morestring=[b]",
  morestring=[s]{>}{<},
  morecomment=[s]{<?}{?>},
  stringstyle=\color{black},
  identifierstyle=\color{Blue},
  keywordstyle=\color{RoyalBlue}
}
\usepackage[inline]{enumitem}
\setlist[enumerate]{leftmargin=*,topsep=0pt,partopsep=0pt}%
\setlist[itemize]{leftmargin=*,topsep=0pt,partopsep=0pt}%

\begin{document}

\date{}
\title{\thetitle}

\author{
	Abdessalam Elhabbash,\textsuperscript{$\spadesuit$} 
	Assylbek Jumagaliyev,\textsuperscript{$\star$}\\
	Gordon S. Blair,\textsuperscript{$\spadesuit$} 
	and Yehia Elkhatib\textsuperscript{$\spadesuit$}\\[4mm]
    \small {$\spadesuit$} School of Computing and Communications, Lancaster University, United Kingdom\\
    \small {$\star$} Z\"uhlke Engineering Ltd., UK\\
    \normalsize Email: a.elhabbash@lancaster.ac.uk\\[4mm]
    \textbf{\textcolor{red}{This is a pre-print}}\\
    \textbf{\textcolor{red}{The final version is available on ACM DL}}\\
}

\maketitle

\thispagestyle{fancy} 

\begin{abstract}
Cloud modelling languages (CMLs) are designed to assist customers in tackling the diversity of services in the cloud market. While many CMLs have been proposed in the literature, they lack practical support for automating the selection of services based on the specific service level objectives of a customer's application. We put forward \name, a novel and generative CML to capture service level requirements and, subsequently, to select the services to honour customer requirements and generate the deployment code appropriate to these services. We present the architectural design of \name and the associated broker that realises the deployment operations. We rigorously evaluate \name using a mixed methods approach. First, we exploit an experimental case study with a group of researchers and developers using a real-world cloud application. We also assess overheads through an exhaustive set of empirical scalability tests. Through expressing the levels of gained productivity and experienced usability, we highlight \name's profound potential in enabling user-centric cloud brokers. We also discuss limitations as application requirements grow.
\end{abstract}

\section{Introduction}
\label{sec:intro}

The growth of the cloud market poses a challenge to its customers who are already overwhelmed with a wide choice of services~\cite{CSC2016}. The scale as well as heterogeneity of the range of offerings and their real time performance variation are adding more complexity to the decision of cloud service selection~\cite{salama2014novel,Kilcioglu2017}, particularly in multi-cloud applications~\cite{daleel,DBLP:journals/corr/abs-1905-02448}. 

Firstly, the \textbf{scale} of cloud services is rapidly growing as more services are offered in the market. A survey of the number of the main cloud providers showed that 198 instance types were offered in 2017 compared to 134 in 2015~\cite{Elhabbash2019brokersurvey}. The number of instance types on offer from Microsoft Azure alone increased more than three times between 2015 and 2017.

Secondly, providers adopt \textbf{heterogeneous} ways to describe instance specifications, pricing, and service level objectives (SLOs)~\cite{Elhabbash2019envisioning}. For instance, Microsoft Azure Cosmos DB and AWS DynamoDB are largely equivalent NoSQL cloud services. They both express the availability SLO\footnote{\textit{Monthly Availability Percentage} for Cosmos DB, and \textit{Monthly Uptime Percentage} for DynamoDB.} in terms of the error rate, \ie the percentage of failed requests during a billing month. However, Cosmos DB error rate is calculated in one-hour intervals whereas DynamoDB measures it in five minute intervals. 

Thirdly, \textbf{unexpected performance} may result in substantial financial losses. Recent analysis of some cloud instances shows that performance levels are inconsistent with the promised offerings~\cite{Zhonghong2012,Hwang2016elastic,Ghrada2018cloudvnf,Baughman2018profiling}. For example, as reported in \cite{daleel}, the performance of a standard workload on an AWS \texttt{c4.xlarge} instance is quite the same as that of \texttt{c4.large} although the former is twice both in specification and cost of the latter.

\begin{figure}[tb]
\centering
\includegraphics[width=0.5\columnwidth]{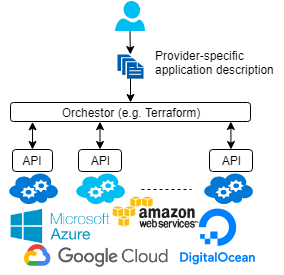}
\caption{The general architecture of an IaC-based system for semi-automated cloud application deployment.}
\label{fig:currentArch}
\end{figure}

In view of the above challenges, the process of manually selecting the optimal service can overwhelm a human decision maker. In order to make it easier for customers to select services and deploy applications, cloud modelling languages (CMLs) were proposed (\eg \cite{Nguyen2011,Guillen2013,Binz2014,Rossini2017,Sandobalin2017,Jumagaliyev2019}). They provide means for composing a high level description of a cloud application topology, then automate their deployment accordingly. Such description, also known as Infrastructure as a Code (IaC), declaratively represents the application architecture, interactions, and the types of required cloud services. An orchestrator can then utilise the IaC model to deploy the application on the cloud, as illustrated in \fig{fig:currentArch}.

There are, however, two main drawbacks with current CMLs. First, they lack the support for automated cloud service selection. Customers first need to identify the service(s) they need, which is challenging due to scale and heterogeneity as discussed. Second, there is a lack of support for modelling SLOs of cloud applications. Customers need to manually compare cloud provides service level agreements (SLAs) in order to select a service based on the required SLOs. A few of the current CMLs support such modelling but through standards that are designed primarily for web service providers to specify their services levels, which is unsuitable for use by cloud customers.

In this paper, we aim to address the aforementioned challenges of scale and heterogeneity in addition to the SLO modelling gap. Our aim is to assist cloud customers in selecting cloud services by achieving interoperability between the provider SLAs and the CMLs. Our approach is to base the selection on provider guarantees regarding service performance. That is, we aim to make selection decisions based on a set of SLOs that are part of the SLAs. However, this requires a customer-oriented language for SLO specification and an engine that realises a SLO-driven selection of cloud services.

Therefore, we propose a design of a new language for SLO modelling, \textit{\name}, that provides a comprehensive syntax for capturing service level requirements, supporting all SLOs currently used by IaaS providers and those specified in industry standards. Through the \name approach, we aim to raise the level of abstraction provided to cloud customers. We adopt a generative language approach whereby customers specify SLOs (\ie develop \name script) for required cloud services regardless of the low level details of those services. Then, the \name script will be translated into deployment code that is utilised by the orchestrator to deploy. 

In addition, we present the architecture of a cloud brokerage system (CBS) that realises the \name approach \cite{Elhabbash2019frame}. The customer will provide an SLO model to the CBS. The CBS will then parse the models, select the cloud services, and generate the deployment model. The broker can also deploy the application on the selected cloud services.

This paper makes the following contributions:
\begin{enumerate}
    \item A novel SLO modelling language, \name, that supports a \emph{comprehensive set} of SLOs for all types of cloud applications and covering all SLAs in the current IaaS market;
    \item An architecture of a brokerage system that utilises \name for cloud service selection; and
    \item A mixed-methods evaluation of the applicability of \name using a real commercial application. Specifically, we assess the added value through a case study experiment with a group of developers of different backgrounds, and we also quantitatively examine the overheads of \name.
\end{enumerate}
\name is available as open source at \url{https://github.com/AbdessalamElhabbash/SLO-ML}.

The rest of the paper is organised as follows. 
\sect{sec:example} motivates the research through a real world application. \sect{sec:sloml} and \sect{sec:arch} present the proposed approach. \sect{sec:eval} evaluates \name through an experimental user study where
real developers are asked to utilise \name for cloud service selection, while \sect{sec:quaneval} evaluates the scalability of the broker architecture. \sect{sec:disc} discusses the findings, limitations, and future work. \sect{sec:rw} comments on related works and \sect{sec:conc} draws conclusions.


\section{Motivating Example}
\label{sec:example}
\begin{figure}[tbh]
\centering
\includegraphics[width=0.4\columnwidth]{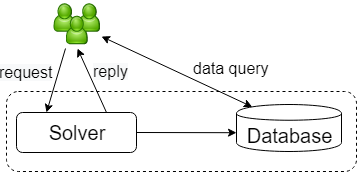}
\caption{The architecture of the SolveEngine application.}
\label{fig:SolveEngineArch}
\end{figure}

SolveEngine\footnote{SolverEngine is a commercial product developed by Satalia, and is available at \url{https://solve.satalia.com/}} is a cloud-based problem-solving service. It aggregates thousands of optimisation algorithms and uses artificial intelligence to choose the best ones to solve optimisation problems. Users format their problems in a supported input format and use the SolveEngine API to call the service. SolveEngine then applies the suitable solvers by using Machine Learning techniques, returning the results in JSON format. SolveEngine is a container-based application that consists of two main components, \textit{Solver} and \textit{Database} (see \fig{fig:SolveEngineArch}). The Solver processes user requests while the Database stores the processing results. Users can also query the Database to obtain certain data of interest.

To deploy such 2-component application in the cloud, the application operator (a cloud customer in this case) needs to manually look into different cloud provider SLAs and assess whether or not they satisfy the application's SLOs. This is a time-consuming and challenging task due to the scale and heterogeneity of service offerings as already highlighted. 
Consider for example an SLO of \textit{Monthly Bandwidth Cost}, which specifies the customer's budget for data transfer between components. The cost calculation depends on several factors such as which provider to use, which service, and in which region. Taking also into account that the cost will be different for different permutations of services, the search space will make the selection decision very challenging for the customer.


\section{\name Design and Concepts}
\label{sec:sloml}

The key aim of \name is to provide a comprehensive syntax to capture all possible SLOs that customers may require to specify service levels of their applications. For this purpose, \name enables customers to specify SLOs for each application component. Moreover, \name supports SLO specification on both single- and multi-cloud deployments.

\begin{figure}[tb]
\centering
\includegraphics[width=0.5\columnwidth]{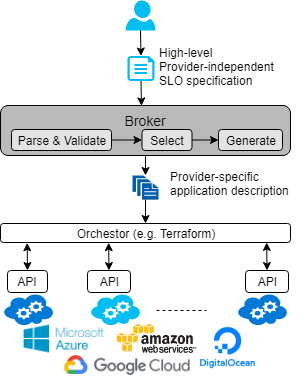}
\caption{The architecture of using \name with the broker. Compared to the general IaC architecture that is common in industry (\fig{fig:currentArch}), the proposed architecture provides much more abstraction and transparency.}
\label{fig:proposedArch}
\end{figure}

\subsection{Design principles}
The design of \name is based on the following principles: 
\begin{enumerate}
    \item \textbf{Customer-oriented.} \name is designed to enable customers to specify their high-level operational requirements in a simple declarative syntax. \name differentiates between two classes of SLOs, namely the service-level SLOs and the application-level SLOs. A service-level SLO represents a quantitative characteristic of the cloud service regardless of the hosted application. Some of the service-level SLOs are specified in the provider SLAs with penalties paid to the customer in case of violation. On the other hand, an application-level SLO represent a quantitative characteristic of the cloud application as perceived by the application client. This kind of SLO cannot be specified in the SLAs as it depends on many aspects of the application such as the application architecture, implementation choices, among other. This implies that the responsibility on satisfying the application-level SLOs is outside of the service provider. This requires an intelligent intermediary system that is able to capture knowledge about the performance of the application when hosted on a certain cloud service and utilise that knowledge to inform the service selection decision.
 
    \item \textbf{Independence}. In order to prevent vendor lock-in, SLO specification needs to be independent of cloud service specification. Furthermore, it needs to be independent of cloud application development technology and implementation details. This is to impose no restrictions on the customer choice of programming models, and to minimise SLO specification changes when adapting the application. In fact, the SLO specifications need to be adapted only when the architecture of the application changes, as the SLOs can be specified per application components.

    \item \textbf{Abstraction}. Customers should be able to specify SLOs regardless of the required type of cloud service, such as SaaS, PaaS, FaaS, \etc.

    \item \textbf{Separation of concerns}. It should be possible to maintain and adapt isolated SLO specification at an application component level. For example, a load-balancing component's SLOs should be separate from those of a data storage element. 
    
    \item \textbf{Mapping SLOs} A high-level SLOs which specified by users should be broken down to low-level ones, and then further mapped to the application component level. For example, the response time of a three-tier application consists of processing time for each layers. 

    \item \textbf{Extensibility}. Extending capability should be simple. In other words, adding a new SLO concept should not require re-engineering of the CML but just adding a human- and machine-readable SLO name along with the appropriate unit and value type, if needed. Obviously, this requires slightly amending the engine that processes the specified model.
\end{enumerate}

\subsection{Key elements}
\label{subsec:elements}
At this stage of designing \name, we adopt textual syntax to represent SLOs. The main elements of the current syntax are: \texttt{name}, \texttt{type}, \texttt{unit}, \texttt{operator}, \texttt{application}, and \texttt{data\_flow}. 

\begin{itemize}

    \item \texttt{name}: A unique keyword is used to refer to each SLO. The keywords are self-explanatory, making it simple for developers to understand. For example, the keyword \texttt{Response\_Time} is used to refer to the response time SLO. 
    
    \item \texttt{value}: \name supports three types of the SLO values: scalar, interval, and categorical. The scalar type is used to specify a numerical value (\eg availability = 0.99). The interval type is used to specify an upper- and lower-bound of SLO value (\eg response time between 5ms and 10ms). Categorical types provide a higher level of abstraction for SLO value specification. It allows customers to specify a category (\eg low, medium, high) instead of specific values or a predefined range, relieving customers from specifying an exact value in case they are not certain. For example, for memory-intensive application, a customer can specify the category high  for the \texttt{Memory\_Size} SLO. 

    \item \texttt{unit}: \name uses a set of keywords that specify the units of measurement of each SLO. For example, the \texttt{Migration\_Time} SLO is specified using the hours unit. In addition, \name contains rules for unit-to-unit conversion between units of the same kind.

    \item \texttt{operator}: \name defines a set of operators that are used to specify the SLO values. This set includes the operators: less than ($<$), less than or equal ($\leq$), greater than ($>$), greater than or equal ($\geq$), equal ($=$), and in ($in$). For instance, $in$ can be used to indicate that \textit{response\_time} should be in the interval \texttt{[5ms,10ms]}.

    \item \texttt{application}: This is to specify the application-level SLOs.
    
    \item \texttt{data\_flow}: This is to specify the directions of data transfer among the application components, which are used in the service selection phase to calculate the expected data transfer costs.

\end{itemize}

\begin{lstlisting}

{    
    "database_comp": { //component 1 
        "SLOs": [
            //service-level SLO
            {    "unit": "", 
                "name": "Monthly_uptime_percentage",
                "value": "0.9999",
                "operator": ">="
            },
            //service-level SLO
            {    "unit": "GB",
                "name": "Monthly_egress_bandwidth",
                "value": "2000",
                "operator": "<="
            }
        ],
        "config": {
            "type": "database"
        }
    },
    "solver_comp": {//component 2
        ...
    },
    "application": {
        "SLOs": [
            //application-level SLO
            {    "unit": "\$",
                "name": "Monthly_bandwidth_cost",
                "value": "20",
                "operator": "<"
            }
        ]
    },
    "data_flow": [{
        "from": "solver_comp",
        "to": "database_comp"
    }]
}
\end{lstlisting}

\subsection{Grammar}
\begin{sloppypar}
We adopt JSON syntax \cite{RFC4627} for structuring the \name file (.slo) that defines the required SLO. This definition is structure as a \texttt{Map<key,value>} where the \texttt{key} is an application component identifier that is defined in the IaC description, while \texttt{value} is an array of maps representing the SLOs required for that component. Each map is a \texttt{Map<key,value>} where the \texttt{key} is one of the elements described in \sect{subsec:elements} and \texttt{value} is the corresponding value. Listing~1 shows an example of the SLO specification of a cloud application that consists of two components, \textit{database\_comp} and \textit{solver\_comp}. The listing shows that \textit{database\_comp} requires two SLOs, \texttt{Monthly\_uptime\_percentage} and \texttt{Monthly\_egress\_bandwidth} at the service-level. The application also requires the \texttt{Monthly\_bandwidth\_cost} which specifies the budget for data transfer of the application. The \texttt{data\_flow} part shows that data will be transferred from the \textit{solver\_comp} component to the \textit{database\_comp}. The use of an invalid element key,  invalid element value, or invalid SLO unit will produce a parsing error. 
\end{sloppypar}


\section{Broker Architecture}
\label{sec:arch}
We provide an architecture for a cloud broker that realises deployment based on user-provided \name descriptions.

\subsection{Overall approach}
Our approach views the cloud application as a set of components, each of which requires a set of SLOs to be specified. The approach builds on existing approaches of modelling cloud applications such as Terraform HCL\footnote{\url{https://www.terraform.io/docs/configuration/syntax.html}}, TOSCA\footnote{\url{https://www.oasis-open.org/committees/tc_home.php?wg_abbrev=tosca}}, \etc. 
We assume that the customer request consists of SLO model defined using \name and the broker will parse the model, select satisfying services, and then generate the CML deployment code (HCL, TOSCA, \etc.). The broker will then execute the deployment code to deploy the application.

\begin{figure}[tb]
\centering
\includegraphics[width=0.65\columnwidth]{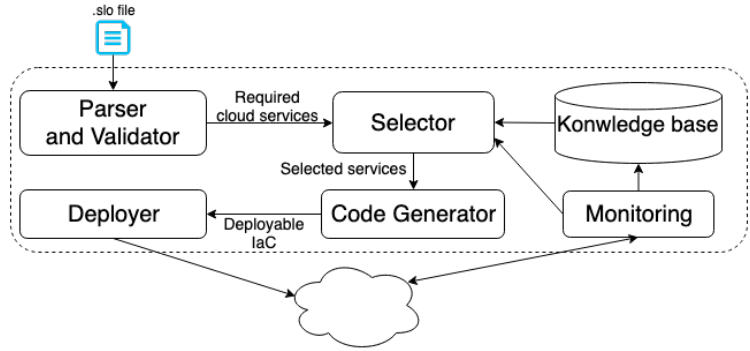}
\caption{The design of the Realisation Engine.}
\label{fig:brokerArch}
\end{figure}

\subsection{Components}
Our proposed broker architecture consists of the following main components, as illustrated in \fig{fig:brokerArch}.

\textbf{Parser and Validator} parses both the SLO and IaC models to extract the required SLOs for each component. The validation intends to evaluate the SLO specification by checking the correctness of the (i) syntax, (ii) units, and (iii) consistency of the configuration. Syntax validation aims at inspecting the syntax for any errors in using \name keywords. Unit validation aims to check for any improper use of units. For instance, the unit \textit{days} cannot be used with the \texttt{Bandwidth} SLO.
Consistency validation ensures that component references in the SLO file correspond to the application components described in the IaC model. 

\textbf{Knowledge Base} is a repository that stores information of the cloud instances such as their type, provider, and the service levels. The Knowledge Base also contains monitoring data that represent the real time performance of the cloud services. 

\textbf{Selector} selects services that match the required SLOs for each component of the application. In its simplest implementation, the selection is based on provider SLAs. More sophisticated implementations may include intelligent selection using monitoring data and consequent predictions of performance. The selection approach adopted in this paper is founded on quantifying the extent to which each service SLO satisfies the required SLO by assigning a utility value to each SLO. These utilities are aggregated to calculate a utility for each cloud service. The utilities are then maximised to select the optimal service(s).

In order to assign utilities for each service SLO, we use the function shown in eq.~(\ref{eqn:sloUtil}), which are adapted from a utility model for quantifying volunteer services~\cite{Elhabbash2014utilitymodel}. The function assigns a minimum utility of $0$ to SLOs that satisfy the corresponding required SLO. The service SLOs that do not satisfy the required one receive a utility of $-1$. The SLOs utilities are then summed up using eq.~(\ref{eqn:sumUtil}) to calculate the service utility. For each combination of services, an application-level utility is calculated using eq.~(\ref{eqn:aggUtil}), where corresponding SLOs are aggregated using suitable  aggregation functions (\eg \textit{sum} for cost and \textit{min} for availability). The application-level utilities are then maximised to select the optimal service(s).

\begin{equation}
\label{eqn:sloUtil}
\small
U_i(S_j) = 
\begin{cases}
1 - e^{SLO_r - SLO_{ji}} , \text{if } SLO_{ji} \ge SLO_r\\
-1, \text{\space otherwise} \\
\end{cases}
\end{equation}
where $SLO_{ji}$ is the $i$th SLO  of service $j$, $SLO_r$ is the corresponding required SLO, and $U_i(S_j)$ is the utility of $SLO_{ji}$. 
\begin{equation}
\label{eqn:sumUtil}
\small
U(S_j) = \sum_{i=1}^{n} U_i(S_j)
\end{equation}
where $U(S_j)$ is the utility of service $j$.

\begin{equation}
\label{eqn:aggUtil}
\small
U_i(comb) = 
\begin{cases}
1 - e^{APP_r - AGG(SLO_i)} , \text{if } AGG(SLO_i) \ge APP_r\\
-1, \text{\space otherwise} \\
\end{cases}
\end{equation}
where $AGG(SLO_i)$ is the aggregate of $i$th SLO of the services, $APP_r$ is the corresponding required application-level SLO, and $U_i(comb)$ is the utility of the combination of services.

\textbf{IaC Code Generator} generates the deployment code of the application based on the selected instances. This deployment code is readily deployable with default settings of the selected cloud services, but customers can customise it as they wish.

\textbf{Deployer} receives the deployment code and automates the deployment of the application on the selected cloud instances.

\textbf{Monitoring} records the low level performance metrics of the selected cloud services. The collected data are stored in the Knowledge Base. The metrics are then mapped to the high level SLOs. If the mapping results in violation of an SLO, the violation is reported to the selector to re-select new instances and adapt the application accordingly. It is worth mentioning that the details of monitoring and adaptation are out of this paper's scope as we focus on presenting the modelling language and the realisation architecture.  


\section{Qualitative Evaluation}
\label{sec:eval}
In this section, we evaluate \name using an experimental case study. We first present the experiment setup and the selected case study, then comment on the results. 

\subsection{Experiment design}
\textbf{Objectives.} The experiment aims at evaluating users productivity, in terms of the time required to select cloud services, and accuracy, in terms of the optimally of SLO offerings of the selected service. 

\textbf{Strategy.} We compare \name approach against the manual selection of cloud services where users need to manually inspect and compare the service specification and SLA offerings in order to select suitable services for the given use cases. We adopt a controlled experiment approach where participants are given three user cases. The use cases are designed to be simple so that they can conveniently doable by the participants within reasonable experimental time. This controlled experiment strategy design is leveraged to evaluate the interaction of the users with \name. The analysis of this interaction enables the identification of advantages and limits of \name in addition to improvements that can be introduced.

\textbf{Procedure.} The experiment procedure lasts for a maximum of an hour per participant. Each participant is assigned three use cases to select cloud services first manually then by developing and executing \name scripts for each case. All participants performed the same use cases and used the same powerful PC. 
At the end of the experiment, the participant fills a questionnaire about their experience in programming languages, cloud service selection, cloud application deployment and application modelling languages and tools. 
Then, each participant is asked to respond to a simple questionnaire using a 5-point Likert scale, to provide feedback about usability and productivity of \name and things to improve.

\textbf{Task}: Participants were given three simple architectures of cloud applications along with their SLO requirements. Each application consists of one or more components, where every component can be deployed on a cloud service that should satisfy the functional and non-functional requirements. Participants are given a list of services that functionally satisfy the components along with the services' SLAs. They need to find services that match the SLO requirements by:
\begin{enumerate}[label=\roman*)]
    \item following the current approach where provider SLAs are manually inspected to find matching services, and 
    \item writing a \name script to be utilised for automated search.
\end{enumerate}

\textbf{Assistance}: Before the experiment commences, participants are introduced to the relevant SLAs of the considered cloud providers, namely, Amazon Web Services, Microsoft Azure, Google cloud, and RackSpace. They are also introduced to \name with a brief quick-start guide (2-3 minutes) and a sample script. During the experiment, additional guidance is provided to any participant requiring assistance for interacting with either the service SLAs and offerings or with \name.

\textbf{Recruitment}: Participants were recruited from Computer Science researchers and students at Lancaster University, as well as from software developers at local startups and incubators. An incentive for participation was offered in the form of an online shopping voucher (value of \pounds10). Overall, 20 participants with varying expertise levels in programming and cloud systems were recruited. These were broken down as 8 researchers, 8 graduate students, and 4 professional developers. Further, 11 of them self-reported high experience (above 5 of a scale from 1 to 7) in JSON, 4 with medium experience (3-4) and 5 with low experience (1-2). Regarding expertise in programming, 8 reported high programming experience (more than 7 years), 5 of medium experience (4-6 years), and 7 with low experience (3 or less years). Finally, 8 self-reported knowledge of cloud application deployment and/or cloud service selection, with AWS and Google Cloud being the most used providers.

\subsection{The SolveEngine case study}
We exploit the SolveEngine application (introduced in \sect{sec:example}) as a real-world case study. We ask the participants to use it under the following three experimental use cases:

\subsubsection{Single component:} In this case, the SolveEngine application is to be deployed on a hybrid cloud where the Solver component is hosted locally whereas the database component is deployed on a cloud service. The customer needs to select a cloud database service to host the database component. Case 1 in table~\ref{tbl:SolveEngineSLOs} lists the required SLOs of the database components.

\subsubsection{Case 2} In this case both components need to be hosted on the cloud. The customer needs to select a cloud database service to host the \textit{database} component and a compute service to host the \textit{solver} component.

\subsubsection{Case 3} This scenario is similar to the previous one. The difference is that the user has application-level SLOs. The participant needs to ensure the aggregate SLOs of the selected services satisfy the application level SLOs. Table~\ref{tbl:SolveEngineSLOs} shows the bandwidth required between the components and required availability of the application. The customer needs to select a cloud service for each component taking into account the bandwidth budget constraint and needs to aggregate the \texttt{Monthly uptime} of the services. 

\begin{table}[th]
\centering
\caption{SLO requirements of SolveEngine}
\label{tbl:SolveEngineSLOs}
\begin{tabular}{ll}
\toprule
\makecell{Component} & \makecell{SLOs} \\
\midrule
\rowcolor{Gray} \textbf{\emph{Case 1}} & \\
\textbf{Database} & monthly uptime percentage $\ge$ 0.99 \\
& monthly consistency percentage $\ge$ 0.9999\\
& monthly latency attainment percentage $\ge$ 0.9999\\
& monthly throughput percentage $\ge$ 0.9999\\
\midrule
\rowcolor{Gray} \textbf{\emph{Case 2}} & \\
\textbf{Database} & monthly uptime percentage $\ge$ 0.9999 \\
\textbf{Solver} & monthly uptime percentage $\ge$ 0.9999 \\
\midrule
\rowcolor{Gray} \textbf{\emph{Case 3}} & \\
\textbf{Database} & monthly uptime percentage $\ge$ 0.9999 \\
& monthly egress bandwidth $\le$ 2 TB \\
\textbf{Solver} & monthly uptime percentage $\ge$ 0.9999 \\
& monthly egress bandwidth $\le$ 2 TB \\
\textbf{Application} & monthly uptime percentage $\ge$ 0.999\\
& monthly bandwidth cost $\le$ $\$$175\\
\bottomrule
\end{tabular}
\end{table}

\begin{figure*}[tb]
\centering
\includegraphics[width=\textwidth]{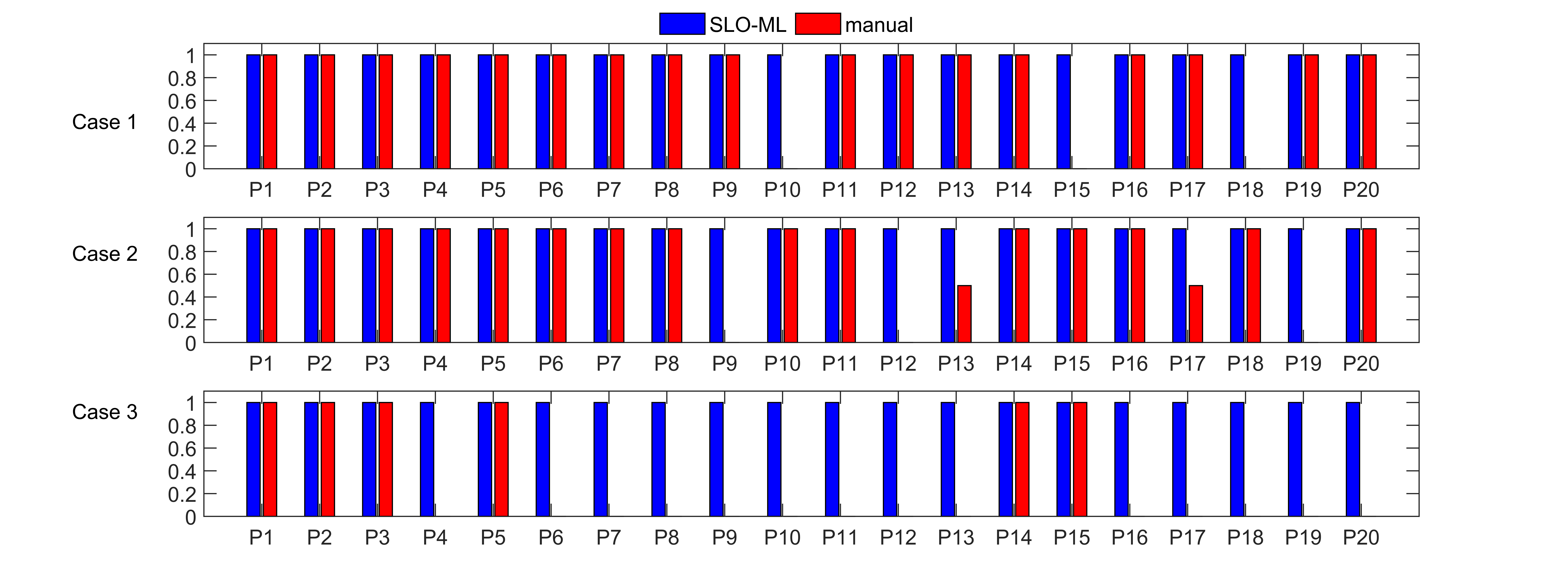}
\caption{Accuracy of service selection of each case using manual and \name approaches.}
\label{fig:accuracy}
\end{figure*}

\begin{figure*}[tb]
\centering
\includegraphics[width=\textwidth]{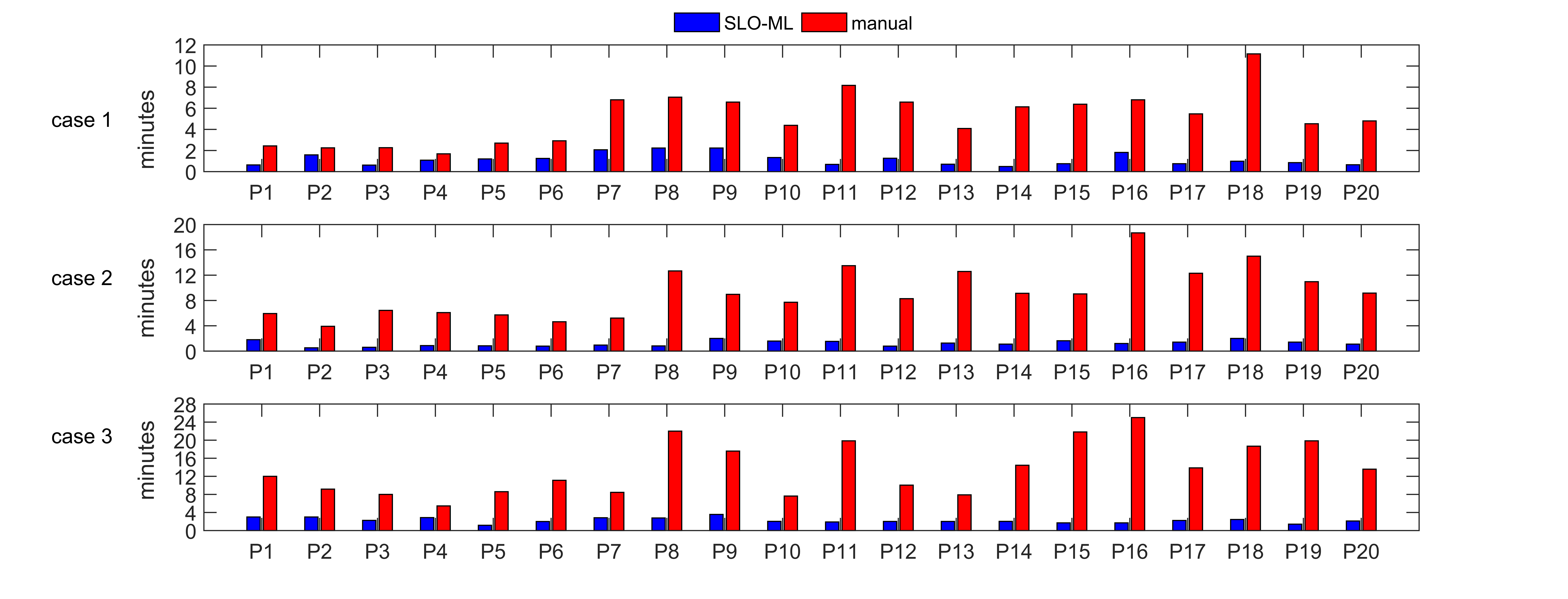}
\caption{The time spent by participants to complete each case using either approaches.}
\label{fig:productivity}
\end{figure*}

\begin{figure}[ht]
\centering
\includegraphics[width=0.75\columnwidth]{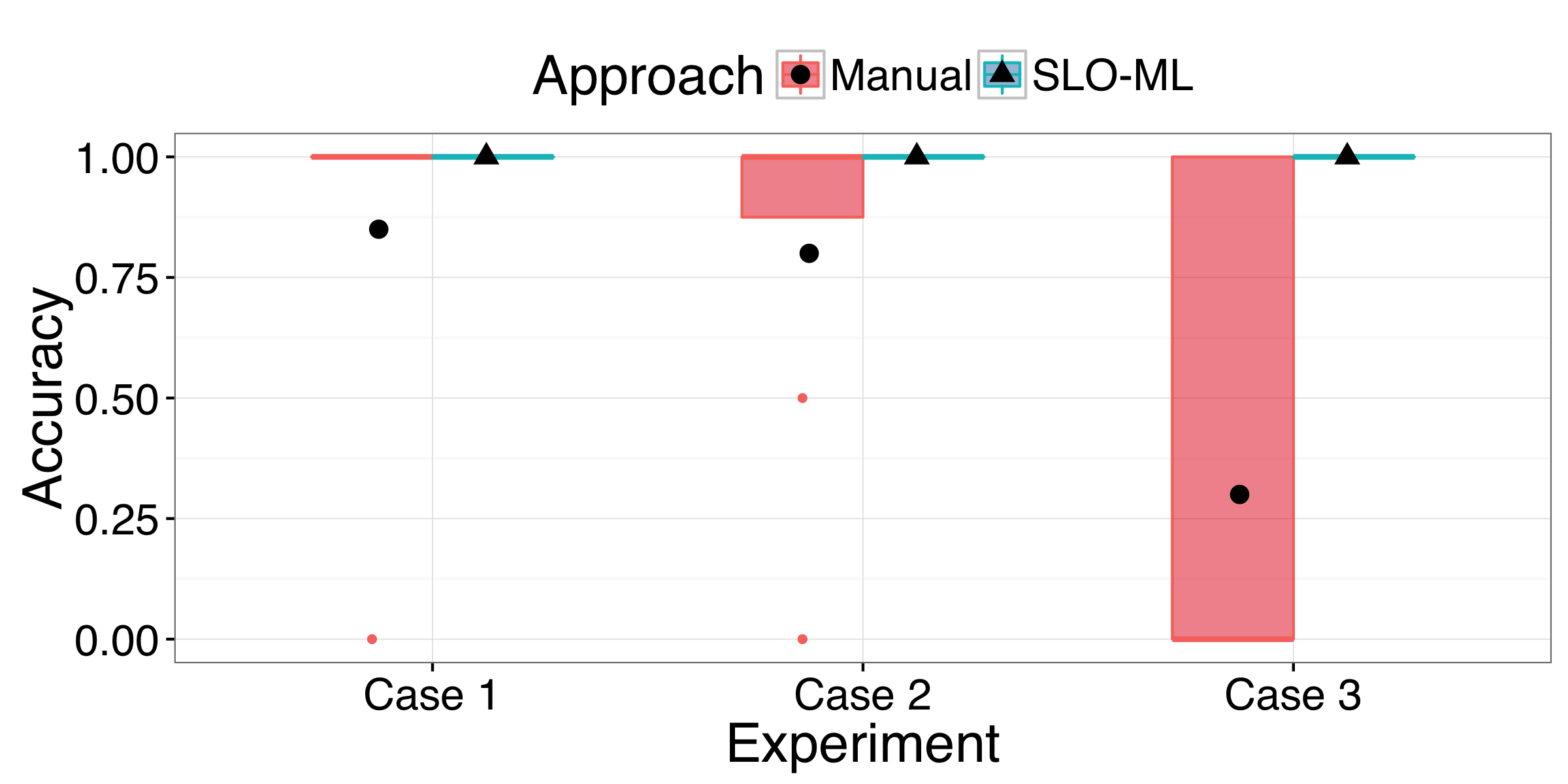}
\caption{Box-plot and mean accuracy of service selection in each experimental case using either approaches. The traditional manual approach creates selection decisions that are further away from the optimum as application SLOs increase in complexity. Meanwhile, \name maintains optimal selection in \textit{all} cases.}
\label{fig:avgAccuracy}
\end{figure}

\begin{figure}[ht]
\centering
\includegraphics[width=0.75\columnwidth]{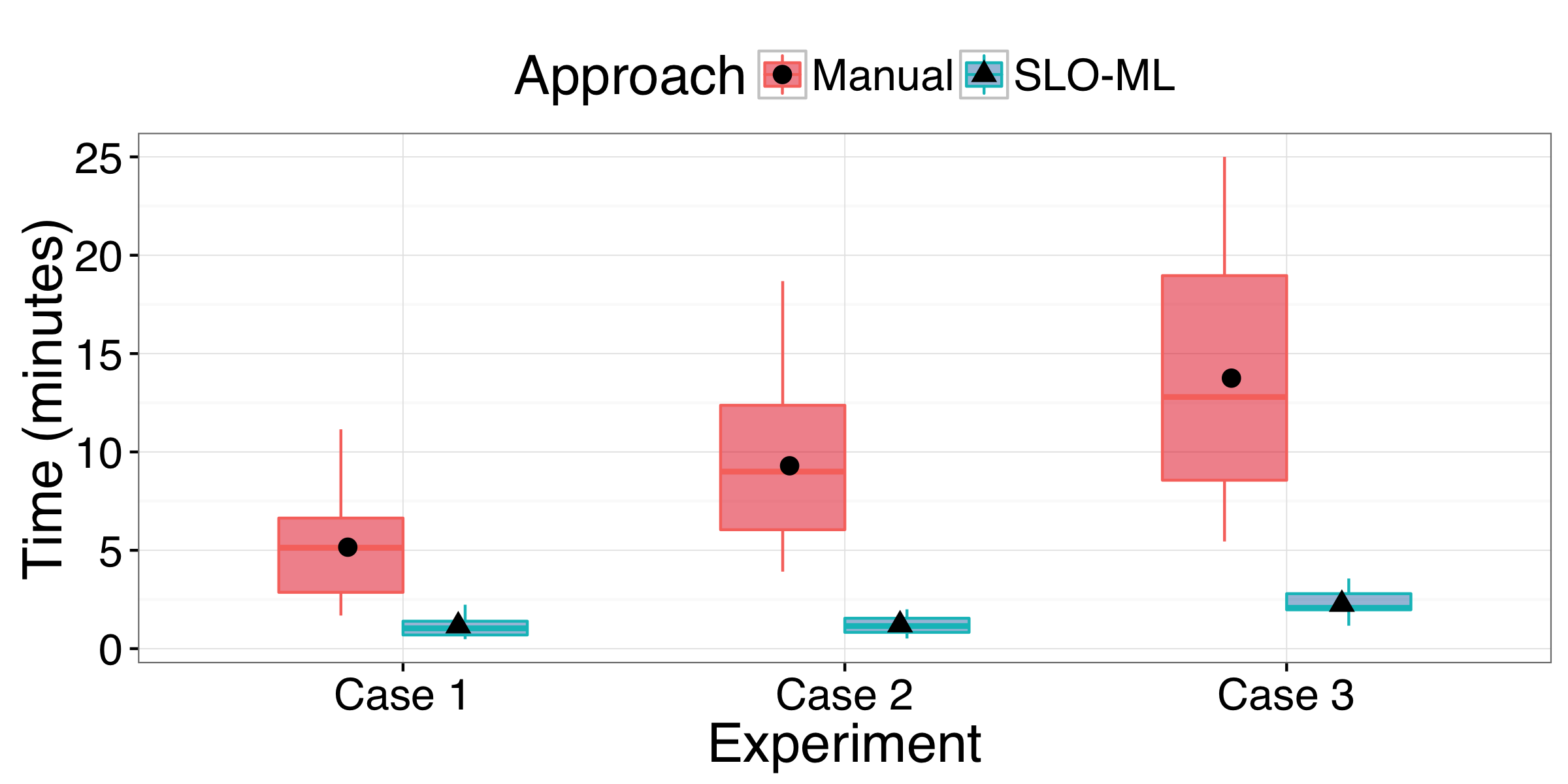}
\caption{Box-plot and mean time spent by participants to complete each case using either approaches. Using the traditional manual approach, developers needed increasingly more time as application complexity grew. In contrast, \name allows them to focus only on SLO specification resulting in significantly reduced time, by a factor of $4.5$--$7.7$x.}
\label{fig:avgTime}
\end{figure}

\subsection{Accuracy results}
The accuracy of the selection is evaluated by calculating the distance between the utility of the selected services and the optimal one. For this, the above utility functions (\sect{sec:arch}) are used to calculate the utility of the selected services. 

\begin{sloppypar}
\fig{fig:accuracy} compares the accuracy of each participant's selection in both the \name and  manual approaches. In all the three use cases, the results demonstrate that \name improves selection accuracy. In case 1, which is the simplest case, most of the participants manually selected the optimal service, Microsoft Azure Cosmos DB. This service is the only one with SLA support of the required SLOs (see table~\ref{tbl:SolveEngineSLOs}). Despite the simplicity of the case, three of the participants (P10, P15, and P18) selected wrong services, \ie services that do not support the required SLOs (namely, \texttt{monthly consistency attainment percentage}, \texttt{monthly latency attainment percentage}, and \texttt{monthly throughput percentage}); hence, these 3 participants scored an accuracy level of 0 for this use case.
\end{sloppypar}

The improvement in accuracy is more notable as the complexity increases in case 2, and even more so in case 3. In order to highlight the improvement, we plot the average accuracy in the three cases in \fig{fig:avgAccuracy}. This figure reveals that there has been a sharp decline in accuracy using the manual approach as the complexity of the case increases. This is in contrast to the \name approach where optimal accuracy is maintained throughout the use cases.

\subsection{Productivity results}

The productivity of participants is evaluated by calculating the time spent to make a decision of service selection. In the case of \name, productivity is calculated as the time spent to develop and execute a valid \name script whereas in the manual approach case it is calculated as the time from the beginning of inspecting the SLA information until deciding on a service. 

\fig{fig:productivity} compares the productivity of each participant in both approaches. In all three use cases, \name significantly reduces selection time. More importantly, the more complex the use case is the more  significant the improvement is. To better demonstrate this, we plot the distributions of the time spent by participants in the three cases in \fig{fig:avgTime}. What can be clearly seen in this figure is the rapid growth of the completion time of the manual approach as the complexity of the case increases. On the other hand, the growth is much slower in the \name case, indicating its ability to assist developers in tackling cloud deployment scenarios of complex selection decisions without a high tax on their time.

\subsection{Exit interview responses}
After the completion of the three cases, the participants were interviewed in order to survey their experience of using \name. They were asked to answer seven question, four of which aimed to assess productivity and three to assess usability. Responses were collected using a 5-point Likert scale with anchors from `Strongly disagree' to `Strongly agree'.

\begin{figure}[tb]
\centering
\includegraphics[width=0.8\columnwidth]{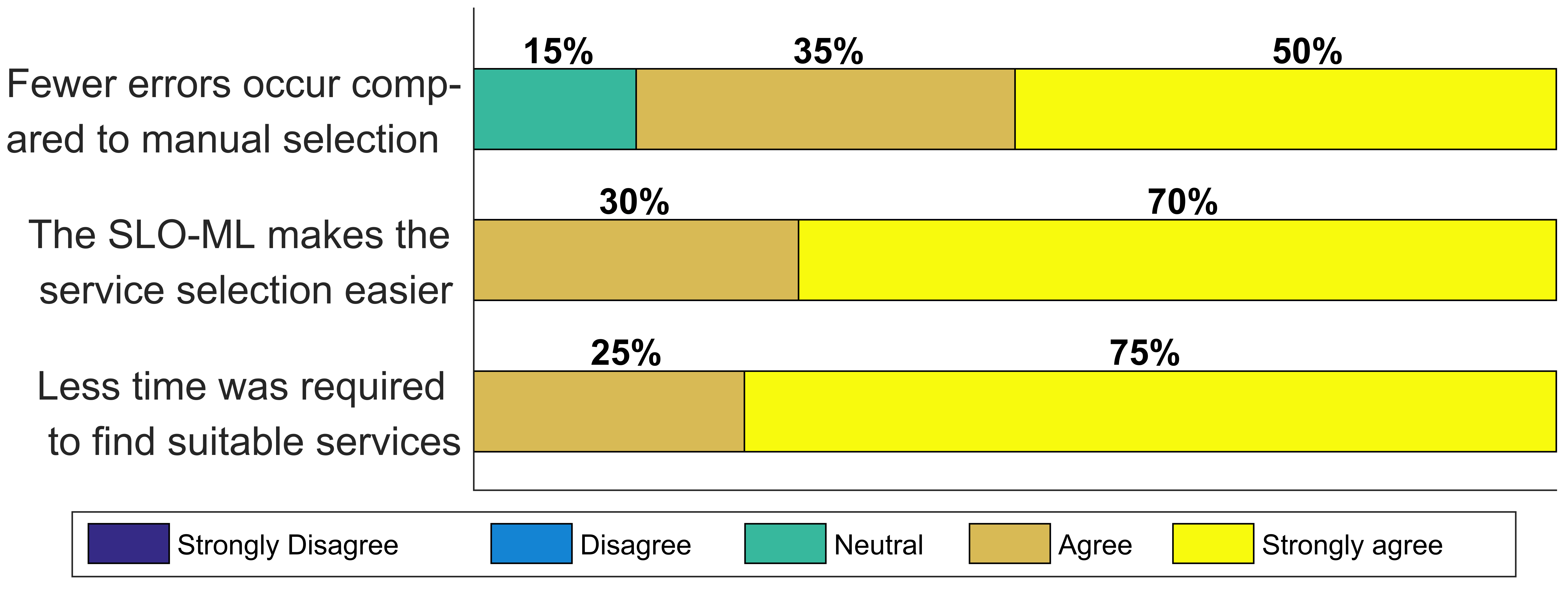}
\caption{Participant feedback on productivity.}
\label{fig:productivityFeedback}
\end{figure}

\begin{figure}[tb]
\centering
\includegraphics[width=0.8\columnwidth]{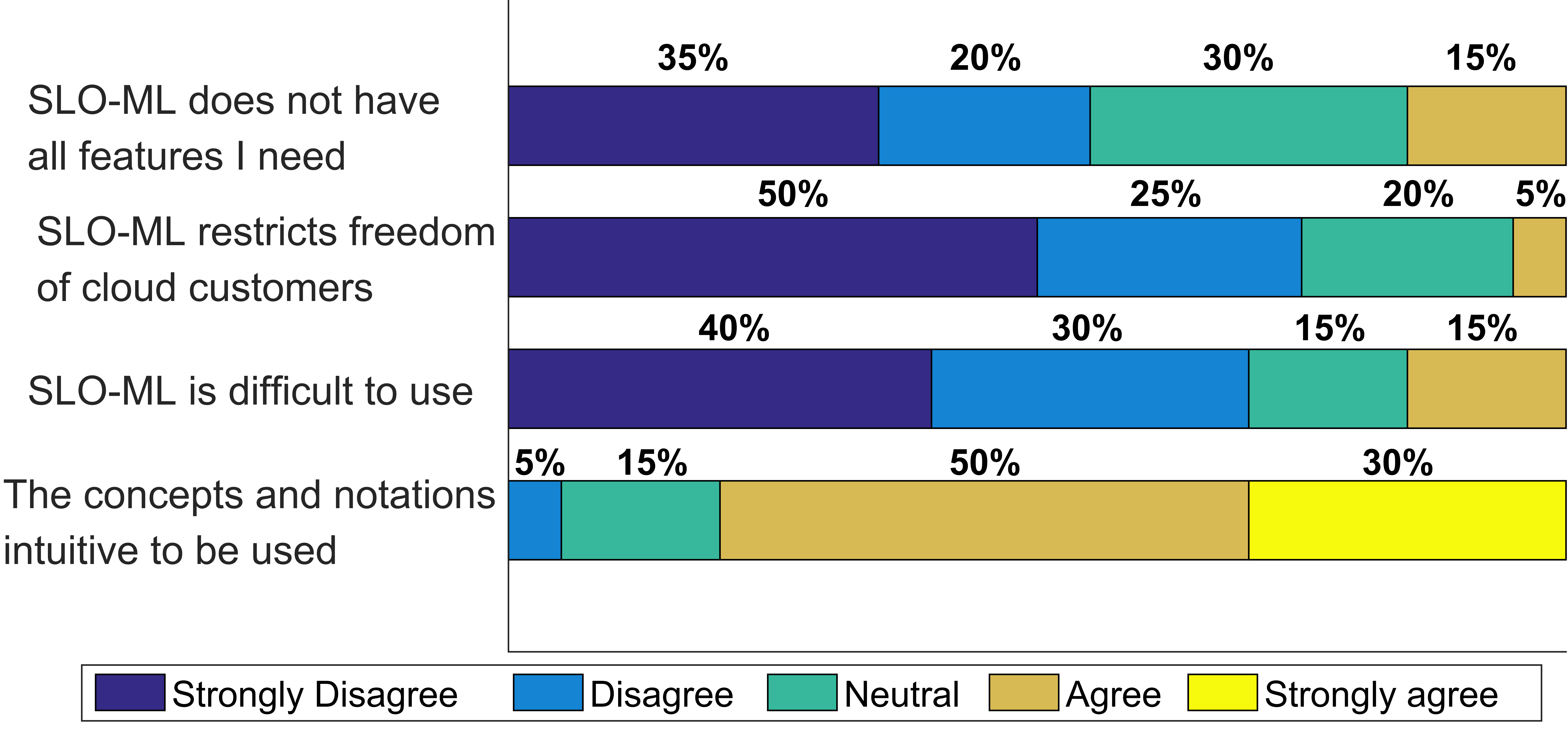}
\caption{Participant feedback on usability.}
\label{fig:usabilityFeedback}
\end{figure}

\textbf{Productivity.} \fig{fig:productivityFeedback} shows the participant feedback on their productivity when using \name. All the participants agreed that less time would be required when using \name especially with complex cases of service selection. All of them also agreed that \name makes service selection easier by automating it as opposed to manual inspection of SLAs. Furthermore, 85\% of the participants agree that \name reduces the possibility of selecting services that do not satisfy the SLO requirements or services that are less optimal. 

\textbf{Usability.} \fig{fig:usabilityFeedback} shows the participant feedback on the usability of \name. The majority of participants found \name and its concepts and notations easy to use and flexible. A few of the participants (15\%) found \name lacks some features they might need, such as support for other SLOs. One participant (5\%) found that \name restricts their freedom as a cloud customer as it does not leave the final decision of selection to them. 


\section{Scalability Evaluation}
\label{sec:quaneval}
We now turn our attention to evaluating the feasibility and overheads of our approach. Specifically, we aim through experimental means to identify the factors that contribute to the end-to-end time of generating the deployment code. For this purpose, we evaluate the time required to parse the \name script, select services, and generate deployment code at different scales. From this, we extract conclusions about the ability of and the requirements for using the \name approach \textit{at scale}. The used platform is an Intel Core~i7 with 16GB RAM running Linux Ubuntu~v16.04 and Java~SE~v1.8.0. Each experiment is repeated 100 times to obtain representative mean values.

\subsection{Parsing time}
This first experiment focuses on measuring the parsing time. This is the time required for analysing the application structure, in terms of the required components and the connections between them, and also the required component- and application-level SLOs. The main dimensions affecting the scalability of parsing are the number of components, the number of SLOs and the degree of connectivity. Specifically, we inspect three degrees of application connectivity that correspond to varying application topologies~\cite{8416514}: 
`low' represents applications such as Riak with a ring-like topology where a component only connects to one or two other components; 
`mid' is analogous to hierarchical hub-and-spoke and other cliquey structures, \eg MongoDB and Ceph; 
while `high' embodies complex applications with highly connected components such as the microservice architecture of the Netflix or Facebook infrastructures.

\begin{figure*}[ht]
\centering
\includegraphics[width=0.99\textwidth]{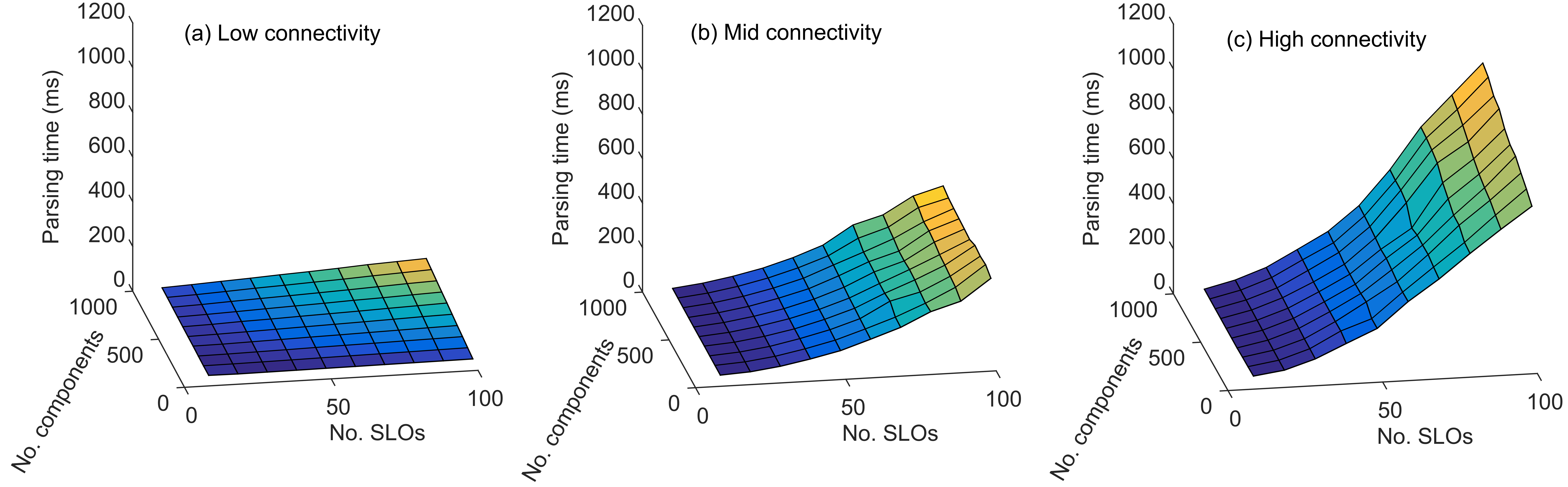}
\caption{Parsing time of \name script with varied scales and connectivity degrees between components.}
\label{fig:parsingTime}
\end{figure*}

\fig{fig:parsingTime} plots the average parsing time in milliseconds. We notice that the parsing time increases with the increase of both the number of SLOs and the number of components. 
In both cases the increase exhibits a linear trend. We notice also that the parsing time increases with the increase in the connectivity degree between the components. However, in all cases the parsing time is practically acceptable, the maximum being $\approx$1s in a very-large scale deployment of  1000 components and 100 SLOs.

\begin{figure*}[tb]
\centering
\includegraphics[width=0.99\textwidth]{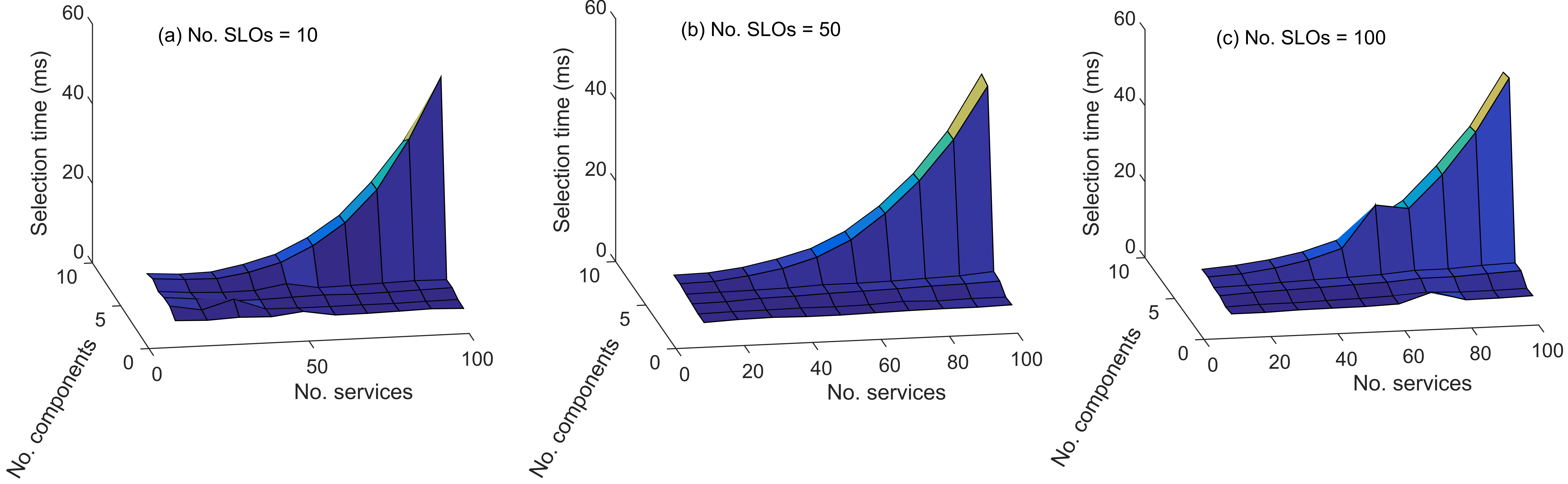}
\caption{\name's selection time as the complexity of an application and the number of its SLOs grow.}
\label{fig:selectionTime}
\end{figure*}

\begin{figure}[ht]
\centering
\includegraphics[width=0.75\columnwidth]{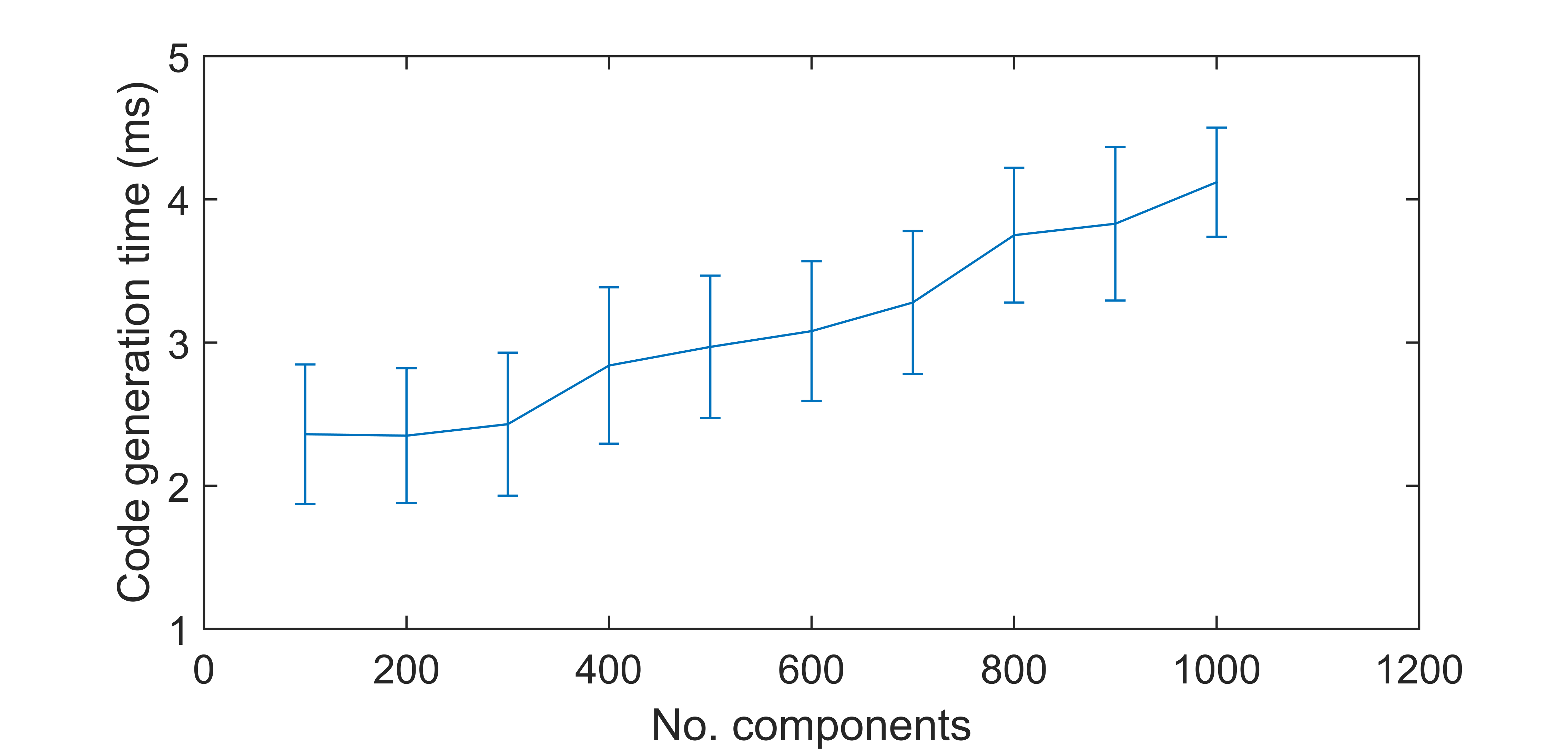}
\caption{\name code generation time for applications with different numbers of components.}
\label{fig:codeGenerationTime}
\end{figure}

\subsection{Deployment code generation time}
The next experiment focuses on assessing the  time required for the step that follows parsing, i.e. generating application deployment code. The generator is written as a Java program that receives information of the selected services and writes to a file CML-specific lines of code. In this paper, we generate code for Terraform deployer, i.e. the generated code is HCL code \footnote{\url{https://www.terraform.io/docs/configuration/syntax.html}}. The only dimension affecting the scalability of code generation is the number of components in an application. We vary this dimension between 100 and 1,000 application components -- see \fig{fig:codeGenerationTime}. We observe a linear trend between code generation time and the number of components. Nevertheless, as the figure depicts, the code generation time is quite insignificant even in the case of a high number of components (\eg $\approx$4ms for 1,000 components), underlining \name's practicality in this regard.

\subsection{Selection time}
The third experiment focuses on evaluating service selection time, defined as the time required to find a single or set of services that satisfy the application SLO requirements. The main dimensions affecting the scalability of selection are the number of components (\ie required services), the number of candidate services for each component, and the number of required SLOs. We vary each of these dimensions, plotting the average selection time in milliseconds in \fig{fig:selectionTime}. Selection time increases significantly with an increase in the number of components, and increases at an exponential rate with the number of candidate services. This implies that the service selection time is the bottleneck of the process especially in the case of a high number of components and/or services. This issue is discussed more in the following section.

\section{Discussion}
\label{sec:disc}
Reflecting on our experiences in developing and evaluating \name, we draw the following observations and concerns.

\subsection{Diversity-induced complexity}
During the experimental case study with SolveEngine, many participants tend to ignore (intentionally or mistakenly) some of the options when selecting the services. For example, some participants made decisions based on a subset of the required SLOs, ignoring the effect of others. This led to the selection of services that either do not satisfy the requirements or do but are less optimal ones. This observation was obvious in case 3 where, perhaps due to its complexity, many participants ignored bandwidth price offerings. We observed that this oversight is due to two reasons. The first reason is the difficulty of finding the relevant SLA documents of the required services. Some participants tended to select services the SLAs of which are easily found. The second reason is the diversity in bandwidth pricing schemes, as several cloud providers charge differently based on region and availability zone. This seemed to confuse some participants as they were not able to determine which offering is best to use, while others did not want to spend time to inspect all of the offerings and opted for randomly selected one. Although complexity is clearly to blame for such behaviour, it clearly can lead to wrong or sub-optimal decision making.
    
Furthermore, some participants made wrong decisions (\ie they selected services that do not satisfy the required SLOs) due to lack of knowledge. They either looked into irrelevant SLAs or they were confounded by the heterogeneous terminology adopted by different providers to express the same SLOs. 

\subsection{Scalability}
Service selection time is the major contributor to the end-to-end time of processing \name script. 
For the worst experimental case illustrated in \fig{fig:selectionTime} where an application has 7 different components and 100 SLOs, the selection time between 100 services is 45 milliseconds which is quite reasonable. Though, this overhead would increase quite rapidly as complexity grows. 
In this paper, we implemented a na\"ive exhaustive search to find the optimal service. However, in the case of applications of a higher scale, more scalable selection algorithms are required. This is beyond the scope of this paper, but luckily the web service selection literature is rich of such selection algorithms \cite{dustdar2005survey}.

\subsection{Experimental validity}
As is common with experimental case study designs, external validity (\ie the ability to generalise the results) is naturally impaired to an extent in order to attain higher internal validity (\ie validating the cause-effect inference). However, by choosing a real-world application that is representative of a range of user-facing cloud applications, we are satisfied that our results are indicative of the significant value added by \name.

\subsection{Future directions}
Currently, \name only caters to SLOs that are supported by provider SLAs. For even more abstract support of application needs, this needs to be extended to include application-specific SLOs that cannot be guaranteed by the provider. An example is the \texttt{Completion time} SLO that specifies a deadline for a certain job. Such extension, however, requires run-time monitoring of the application and continuous assessment of the SLO's satisfaction. This also requires the development of adaptation techniques to adapt service selection in case the current selection fails to satisfy the SLOs. In turn, this also requires the accumulation of knowledge about services and application performance, and the utilisation of such knowledge to predict performance before making the selection decision. These issues are the focus of our ongoing work to extend the \name approach.

\section{Related Work}
\label{sec:rw}

A CML uses modelling concepts to raise the level of abstraction, enabling customers to describe their specific application needs that could then be systematically matched against cloud service offerings. As such, CMLs have been used to design different aspects of cloud application engineering
\cite{Bergmayr2018}. Many CMLs (\eg Blueprint, CAML, CloudDSL, GENTL, CAMEL \cite{Rossini2017}) address the deployment of services and application components to a cloud environment by describing deployment configurations. Meanwhile, other CMLs (\eg CloudMIG, StratusML, TOSCA) deal with the automation of cloud resources provisioning, application migration to the cloud and re-configuration of provisioned cloud services.

A prominent example is the Topology and Orchestration Specification for Cloud Applications (TOSCA)~\cite{Atrey2015:TOSCA}, an OASIS standard for describing the structure of cloud applications (\ie components and relationships) in XML format. 
Similar efforts include GENTL~\cite{Andrikopoulos2014}, 
CloudML-UFPE~\cite{Gonalves2011CloudMLAI}, and CloudML-SINTEF~\cite{BergmayrRFHOSW15}. 

Blueprint~\cite{Nguyen2011} provides concepts for representing service-based applications to facilitate deployment and migration on cloud services. The provided concepts also allow for the representation of different cloud service offerings. 
MULTICLAPP~\cite{Guillen2013} introduces a UML-based profile to model components that can be annotated with deployment information. 
StratusML~\cite{StratusML2015} adopts a similar approach. CAMEL \cite{Rossini2017} can be viewed as a `superset' CML as it integrates and extends existing DSLs. ARGON \cite{Sandobalin2017}, addresses the issue of abstracting the complexity of using CMLs by enabling users to specify infrastructure resources then generating deployment code, similar to the SLO-ML approach. 
CadaML~\cite{jumagaliyev2019lang} is used to manage multi-tenant architecture evolution by transforming an abstract model into the appropriate code for different cloud data storage types.

There are two main shortcomings of the above and other CMLs. First, they require the customer to develop a service-specific IaC model, which means customers need to manually select the cloud services. Such IaC model can be complex to develop from scratch, especially for large-scale applications. Second, they provide limited support for modelling customer SLOs. Instead, they seem to have been designed with a simplistic representation of the provider's perspective not that of the customer. For instance, Blueprint assumes the presence of a marketplace where providers can publish their service descriptions as WS-Policy files. WS-Policy is intended to specify non-functional properties of unary web services, not the complex cloud services customers use today. Furthermore, such assumed marketplace does not exist~\cite{Elkhatib2016crosscloudmap,Elhabbash2019brokersurvey}, so such CMLs are of little practical value in real-world deployments. SLO-ML addresses the above shortcomings and raises the level of abstraction provided to cloud customers.

\section{Conclusion}
\label{sec:conc}
We presented \name, the first cloud modelling language to automate the selection of cloud services to satisfy customer service level objectives (SLOs) and generate the appropriate deployment code. \name is specifically designed to capture a wide range of SLOs from customers. Our findings from an experimental case study suggest that the raised level of abstraction provided by \name results in significant improvements in both developer productivity and optimal service selection. We also identified the limitations of \name, which poses a number of research questions for future work in this area.

 \section*{Acknowledgments}
This work was supported by the Adaptive Brokerage for the Cloud (ABC) project, UK EPSRC grant EP/R010889/1. We thank our project partner Satalia for access to their SolveEngine software. 

\bibliographystyle{ieeetr}
\bibliography{ref,models}

\end{document}